\documentclass[%
 reprint,
showpacs,
 amsmath,amssymb,
 aps,
 prl,
]{revtex4-1}

\usepackage{graphicx}% Include figure files
\usepackage{dcolumn}% Align table columns on decimal point
\usepackage{bm}% bold math
\usepackage{color}
\usepackage{ulem}

\begin{document}

\preprint{APS/123-QED}

\title{Observation of thermal equilibrium in capillary wave turbulence}

\author{Guillaume Michel}
\email{email: guillaume.michel@ens.fr}
\author{Fran\c{c}ois P\'etr\'elis}%
\author{St\'ephan Fauve}%
 \affiliation{%
Laboratoire de Physique Statistique, \'Ecole Normale Sup\'erieure, CNRS, Universit\'e P. et M. Curie, Universit\'e Paris Diderot, Paris, France}
\date{\today}
\begin{abstract}
We consider capillary wave turbulence at scales larger than the forcing one. At such scales, our measurements show that the surface waves dynamics is the one of a thermal equilibrium state in which the effective temperature is related to the injected power. We characterize this evolution with a scaling law and report the statistical properties of the large-scale surface elevation depending on this effective temperature.
\end{abstract}

\maketitle

\paragraph{Introduction.---}One of the striking features of turbulence is the very wide range of time-scales over which the dynamics evolves, from less than $10^{-4}\mathrm{s}$ up to several days in laboratory experiments (see \textit{e.g.} \cite{Herault2015}) and way more in astrophysical systems ($\sim 200 \mathrm{kyr}$ for the mean time between reversals of the earth magnetic field, triggered by turbulent fluctuations). Until recently, these statistical properties of the flow were either related to the transfer of a quadratic conserved quantity through scales, the so-called Kolmogorov cascade and equivalents, or to the chaotic dynamics of a structure (\textit{e.g.} reversal of a large-scale flow \cite{Herault2015}, motion of a shear layer in a von K\'arm\'an swirling flow \cite{Ravelet2008}, \textit{etc.}).  In the case of an homogeneous flow conserving a single quadratic invariant, namely energy, another scenario involving a thermal equilibrium state was referred to in order to describe scales larger than the forcing one (see \cite{Frisch1995}, p209). This has been recently confirmed by numerical simulations \cite{Dallas2015} but is so far in lack of experimental evidence.

The very same description applies to surface wave turbulence, which consists in the statistical study of many interacting surface waves \cite{Nazarenko2011}, an example of interest being the ocean. Again, most studies have focused on direct or inverse cascades (see \textit{e.g.} \cite{Falcon2007,Deike2011}) or on the dynamics of localized structures (rogue waves \cite{Chabchoud2011}, Faraday patterns \cite{Tufillaro1989}, solitons, \textit{etc.}). If energy is the only quadratic conserved quantity, weak wave turbulence theory predicts its transfer to small scales through a direct cascade (similar to the Kolmogorov one), whereas it has been predicted that large scales are in thermal equilibrium \cite{Balkovsky1995}. Capillary waves are the simplest experimental system concerned by these predictions and experiments were carried out in a thin layer of liquid helium to study these large scales \cite{Abdurakhimov2015}. However, because of sizable dissipation at these scales, in place of the expected equilibrium state was another energy cascade. Such a regime of bidirectional energy cascade was later theoretically described in the framework of wave turbulence \cite{Lvov2015}.

In this letter, we report the observation of a thermal equilibrium regime in capillary wave turbulence, we show that this out-of-equilibrium system can be fully described with an equilibrium statistics in the low frequency range. This is achieved by using a large depth of fluid with low kinematic viscosity (mercury), hence reducing both viscous drag at the bottom and bulk dissipation. We describe the statistical properties of this state (energy spectrum and probability distribution function of the surface elevation) and show how the effective temperature can be related to the injected power sustaining the out-of-equilibrium state.

\paragraph{Equilibrium and out-of-equilibrium power spectra.---}Before describing the experiment, we briefly sum up theoretical results. In the limit of large depth, surface waves follow the dispersion relation
\begin{equation} \label{disp}
\omega^2 = g k + \left( \frac{\sigma}{\rho} \right) k^3,
\end{equation}
where $\omega$ is the angular frequency, $k$ is the wave number, $g$ is the acceleration of gravity, $\rho$ is the density of the fluid and $\sigma$ is the surface tension. The gravity and capillary terms are equal at a frequency $f_\mathrm{g.c.} = (2\pi)^{-1}(4\rho g^3/\sigma )^{1/4}$ and we thereafter consider waves at larger frequencies, \textit{i.e.} capillary waves.

If the system is in thermodynamic equilibrium, the isotropic energy spectral density per unit density and per unit surface $e_k^{(1D)}$ is given by
\begin{equation} \label{energy_th}
e_k^{(1D)} \mathrm{d}k = k_\mathrm{B}T \times \frac{2\pi k \mathrm{d}k}{(2\pi / L)^2} \times \frac{1}{\rho L^2} \Longrightarrow e_k^{(1D)}= \frac{k_\mathrm{B}T k}{2\pi \rho}.
\end{equation}
$e_k^{(1D)}$ can be related to the power spectrum density of the surface elevation \textit{via} $S_\eta (k) =(\sigma / \rho)^{-1} k^{-2 }e_k^{(1D)} $,
\begin{equation}
S_\eta(k) = \left( \frac{\sigma}{\rho} \right)^{-1} \frac{ k_\mathrm{B}T}{2\pi \rho} k^{-1}.
\end{equation}
Using \eqref{disp}, the power spectrum density in the frequency domain $S_\eta(f) = 2\pi (\mathrm{d}k/\mathrm{d}\omega) S_\eta(k)$ is
\begin{equation}
S_\eta(f)=\frac{k_\mathrm{B}T}{3\sigma \pi f}. \label{eqth}
\end{equation}

Thermal equilibrium of capillary waves with no external forcing (\textit{i.e.} in which $T$ is the room temperature) has been measured by light scattering in the sixties, for instance to investigate the validity of \eqref{disp} at high frequencies \cite{Katyl1968} or the behavior of surface tension in the vicinity of the liquid-vapor critical point \cite{Bouchiat1969}.

\begin{figure}[t]
    \begin{center} \includegraphics[width=7.5cm]{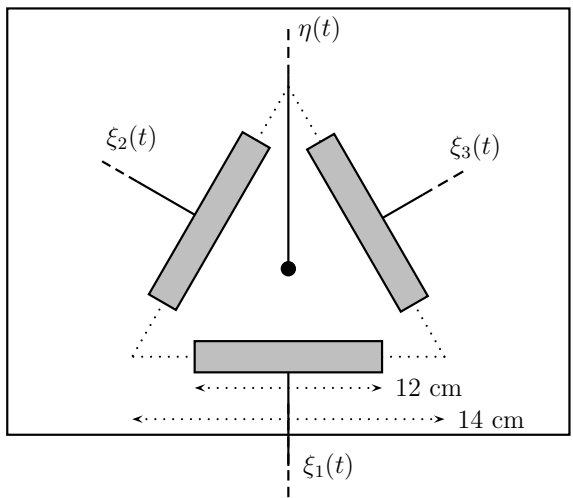} \end{center}
        \vspace*{-.5cm}

    \caption{Experimental setup, consisting in three wave-makers and a capacitive height sensor in a rectangular vessel.}
   \label{manip}
 \end{figure}

On the other hand, if an energy input at a single frequency $f_\mathrm{inj}$ is considered, the system is no longer in thermal equilibrium but eventually reaches a non-equilibrium steady-state. It is then expected to display a direct energy cascade characterized by \cite{Zakharov1967}
\begin{equation}
S_\eta(f) \sim  \epsilon^{1/2}\left(\frac{\sigma}{\rho}\right)^{1/6}f^{-17/6},
\label{weak_turbulence}
\end{equation}
where $\epsilon$ is the mean energy flux through scales (per unit mass and surface). Wave turbulence theory predicts that even if such a system is very far from equilibrium, \eqref{eqth}  should still be observed in the range $f_\mathrm{g.c.}<f<f_\mathrm{inj}$ while \eqref{weak_turbulence} describes frequencies from $f_\mathrm{inj}$ up to a dissipative scale. Matching \eqref{eqth} and \eqref{weak_turbulence} at the single energy input scale of frequency $f_\mathrm{inj}$ gives
\begin{equation} \label{temperature_vs_p}
k_\mathrm{B}T \sim \epsilon^{1/2}\sigma^{7/6}\rho^{-1/6}f_\mathrm{inj}^{-11/6},
\end{equation}
that was obtained within the weak turbulence framework in \cite{Balkovsky1995}. Note that all these results are derived in the absence of gravity waves, which follow a different path since another quadratic quantity called wave action is conserved. From a practical point of view, they are expected to hold as long as the energy density at frequency $f_\mathrm{g.c.}$ is small enough (\eqref{disp} and \eqref{energy_th} show that energy density increases with frequency in the thermal range). Finally, we point out that the derivation of \eqref{eqth} reduces surface waves to harmonic oscillators: at very high temperatures, non-linear corrections are expected.

\paragraph{Experimental setup and results.---} The experimental setup consists of a rectangular plastic vessel ($225\times 180\times 45 \mathrm{mm}$) filled with mercury up to 30 mm. For this fluid, the density is $\rho = 13.5\times 10^3~ \mathrm{kg}/\mathrm{m}^{3}$, the surface tension is $\sigma = 0.485~\mathrm{N}/\mathrm{m}$ and the kinematic viscosity is $\nu = 1.15 \times 10^{-7}~ \mathrm{m}^2/\mathrm{s}$. This parameters give 
$f_\mathrm{g.c.}\simeq 16 \mathrm{Hz}$ and a corresponding wavelength of $12~\mathrm{mm}$, making drag friction at the bottom irrelevant. To get reproducible results, care has been taken to keep the  surface free from pollution: similarly to clean water in which damping increases during approximately one hour until surface gets fully contaminated \cite{VanDorn1966}, oxydation affects mercury. This effect can be significantly reduced by cleaning the surface after every acquisition (every ten minutes).

Three wave-makers (oscillating paddles of size $120\times 80 \times 4 \mathrm{mm}$ plunged 10 mm below the free surface) are placed regularly around a home-made capacitive height sensor (see Fig. \ref{manip}). This size of cavity has been chosen so that the frequencies of the first eigenmodes are a few hertz, thus reducing substantially the gravity range. The diameter of the wire, 0.35mm, imposes a high-frequency cut-off of a few hundred Hertz. Each wave-maker is driven by a Br\"uel \& Kj\ae r 4810 shaker and its motion $\xi_i(t)$ ($i=1,2,3$) is tracked with a Br\"uel \& Kj\ae r 4393 accelerometer. Functions $\xi_i(t)$ are different realizations of a random noise excitation supplied by a function generator (Agilent 33500B) and selected in a frequency range $f_\mathrm{inj} - 200\mathrm{Hz}$ by a SR 650 filter. Several forcing amplitudes and frequencies $f_\mathrm{inj}$ have been considered, the latter varying from 50 Hz to 100 Hz. After each acquisition, the motions of the wave-makers have been checked to be of similar power spectra, localized in the range $f_\mathrm{inj} - 200\mathrm{Hz}$. Finally, the wave height $\eta$ and the accelerations $\ddot{\xi}_i$ are recorded with a NI acquisition card and 50 Hz electrical noise is filtered out and not considered in the following data treatments. 

 \begin{figure}[t]
    \begin{center} \includegraphics[width=8.6cm] {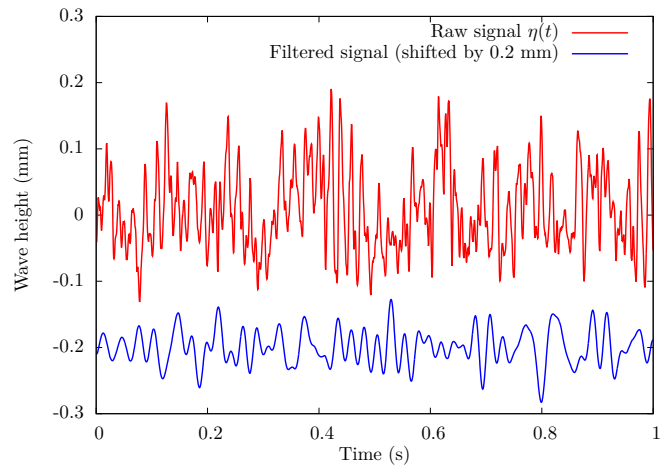} 
    \end{center}
    \vspace*{-.5cm}
    \caption{(Color online) Typical elevation signal and filtered data ($f_\mathrm{inj}=50\mathrm{Hz}$).}
   \label{samples}
 \end{figure}

A typical elevation signal is shown in Fig. \ref{samples}, together with the same sample filtered in the range $f_\mathrm{g.c.}$ to $0.85 f_\mathrm{inj}$: we observe that a sizable amount of energy is located at frequencies lower than the forcing ones. In all the reported experiments, the standard deviation of the height signal $\sigma_\eta = \sqrt{\langle \eta^2(t) \rangle } $ ranges from $0.02\mathrm{mm}$ to $0.1\mathrm{mm}$, corresponding to significant typical steepnesses since the forcing is at high frequencies ($k_\mathrm{inj} \sigma_\eta$ is between 0.04 and 0.2). Moreover, we checked that all the results presented here do not strongly rely on the specific value of the cutoff frequency $0.85f_\mathrm{inj}$.

 \begin{figure}[t]
    \begin{center} \includegraphics[width=8.6cm] {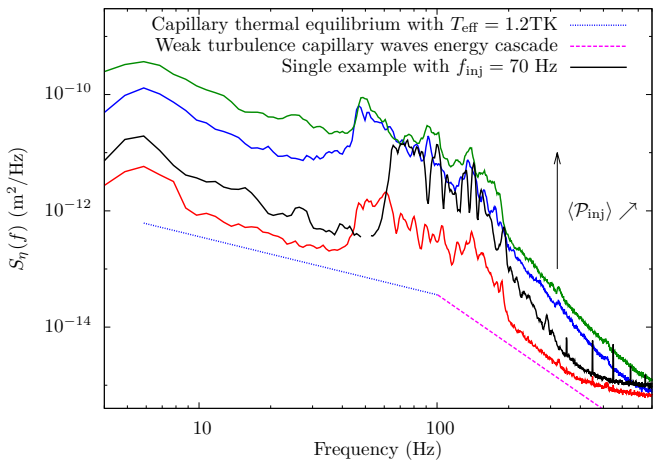} 
    \end{center}
        \vspace*{-.5cm}
    \caption{(Color online) Power spectra of the wave height for three forcing amplitudes with the same bandwidth (50 to 200 Hz) and one from 70 to 200 Hz. A theoretical capillary thermal equilibrium state and direct energy cascade are shown in dotted and dashed lines. $\langle \mathcal{P}_\mathrm{inj}\rangle$ is the mean injected power.}
   \label{spectres}
 \end{figure}

Four power spectra of the wave height $\eta$ are displayed in Fig. \ref{spectres} along with the theoretical ones of a thermal equilibrium state at temperature $T_\mathrm{eff}=1,2.10^{12}\mathrm{K}$ and of a direct energy cascade. They were obtained with $f_\mathrm{inj}=50 \mathrm{Hz}$ and different forcing amplitudes, or with $f_\mathrm{inj}=70 \mathrm{Hz}$. From $f_\mathrm{g.c.}$ to the forcing frequency $f_\mathrm{inj}$, they display a $f^{-1}$ power-law, characteristic of a thermal equilibrium state. In some spectra the $f^{-1}$ slope develops below $f_\mathrm{g.c.}$: similarly to the gravity-capillary transition observed in surface wave turbulence forced at low frequencies, $f_\mathrm{g.c.}$ is only an indicator of the crossover, that may reasonably differ from it (see \textit{e.g.} \cite{Falcon2007}). From 200 Hz to a dissipative scale, another self-similar regime is observed. Because the energy flux is in practice not conserved during this energy cascade \cite{Deike2014, Pan2015} and as a consequence of the high-frequency cut-off of the measurement device, steeper power-laws than the theoretical scaling \eqref{weak_turbulence} are observed. This cascade has been the subject of an extensive literature over the past two decades and will not be further investigated here (see for instance \cite{Henry2000, Falcon2007, Falcon2009, Falcon2011, Deike2014B, Pan2015} and references therein).

For each of these spectrum, we compute the effective temperature by integrating the height power spectrum in the thermal range: 
\begin{equation}
T_\mathrm{eff}= \frac{3\pi \sigma }{k_\mathrm{B}} \frac{\int_{f_\mathrm{g.c.}}^{0.85 f_\mathrm{inj}} S_\eta(f) \mathrm{d}f}{\ln (0.85 f_\mathrm{inj}/f_\mathrm{g.c.})}
\label{Teff}
\end{equation}

\begin{figure}[t]
    \begin{center} \includegraphics[width=8.6cm] {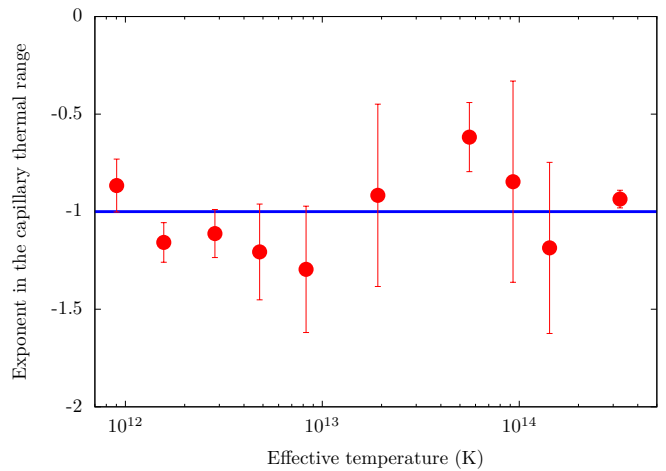} \end{center}    \vspace*{-.5cm}

\caption{Exponent of the large-scale range ($f_\mathrm{g.c.}$ to $0.85 f_\mathrm{inj}$) as a function of the effective temperature.}
\label{expvsT}
\end{figure}

The power spectrum $S_\eta (f)$ is fitted by a power-law $\mathrm{cst} \times f^\alpha$ in the same frequency range ($f_\mathrm{g.c.}$ to $0.85 f_\mathrm{inj}$), and values of $\alpha$  are reported in Fig. \ref{expvsT}. They stand close to -1, as expected for a thermal equilibrium range.

\begin{figure}[t]
    \begin{center} \includegraphics[width=8.6cm] {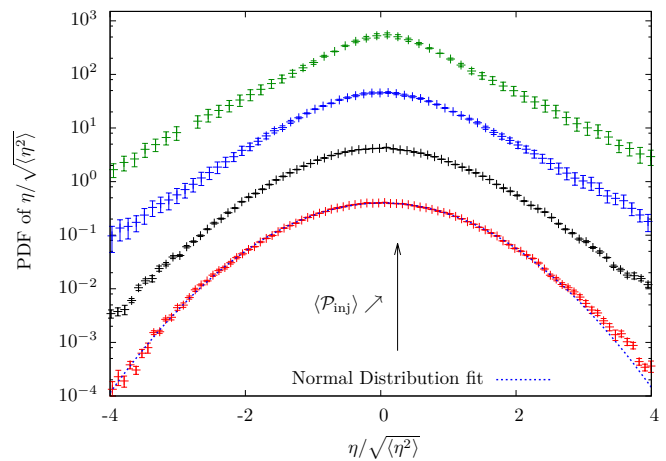} \end{center}    \vspace*{-.5cm}

\caption{(Color online) Probability density functions of the large-scale wave height for the spectra reported in Fig \ref{expvsT} (same color-key). Data and errorbars are multiplied by $\times 1$, $\times 10$, $\times 100$ and $\times 1000$ for clarity.} A fit by a normal distribution is shown in dotted line.
\label{PDF}
\end{figure}

\begin{figure}
    \begin{center} \includegraphics[width=8.6cm] {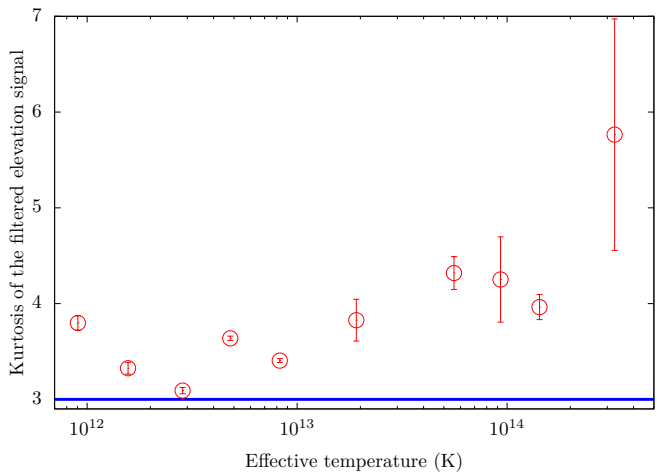} \end{center}    \vspace*{-.5cm}

\caption{Kurtosis of the probability density functions as a function of the effective temperature.}
\label{kurtosis}
\end{figure}

Four probability distribution functions (PDF) of the height signal filtered in the same range ($f_\mathrm{g.c.}$ to $0.85 f_\mathrm{inj}$) and corresponding to the spectra reported in Fig. \ref{spectres} are shown in Fig. \ref{PDF}. The PDF are found to be Gaussian at low forcing and tails turn to exponentials as the injected power increases. To be more precise the kurtosis of the signal, equal to 3 in the case of a normal distribution, is reported in Fig. \ref{kurtosis} and is found to increase with the effective temperature, \textit{i.e.} the steepness of the waves. The same phenomenon was reported in wave turbulence \cite{Falcon2007,Abdurakhimov2015} and ascribed to extreme events as rogue waves \cite{Ruban2006, Shats2010}, that can also take place in a thermal equilibrium state when the steepness of the waves is high enough. Note that departure from a normal law is a sign of sizeable non-linear interactions between modes, as the central limit theorem fails in presence of correlated random variables. As for the other moments of this filtered signal, the standard deviation evolves as the square root of $T_\mathrm{eff}$ (a consequence of \eqref{Teff}) and the skewness is roughly constant and close to 0.20.

The sign of the skewness is related to the shape of the wave-trains: for gravity waves, non-linearities sharpen the crests and flatten the troughs \cite{Stokes1847}, whereas the opposite occurs for capillary waves \cite{Crapper1957}. These phenomena directly reflect on the skewness of the PDF for a random wave field. A positive skewness it routinely observed in experiments of wave turbulence involving gravito-capillary waves \cite{Falcon2007, Falcon2011}. For pure capillary waves, these effects are scarce: a recent numerical simulation of capillary wave turbulence \cite{Deike2014B}, even though involving large wave steepnesses ($\sim 0.3$), has for instance not evidenced any negative skewness. In our experiments, the skewness is positive since the height spectra of the signal filtered between $f_\mathrm{g.c.}$ and $0.85 f_\mathrm{inj}$ is dominated by components at $f_\mathrm{g.c.}$, and vanishes as the low-pass filtering frequency increases. 

\begin{figure}
    \begin{center} \includegraphics[width=8.6cm] {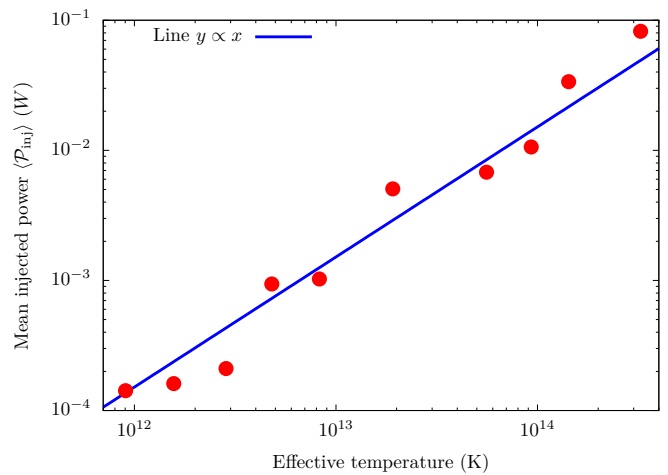} \end{center}    \vspace*{-.5cm}

\caption{Evolution of the mean injected power $\langle \mathcal{P}_\mathrm{inj} \rangle$ with the effective temperature.}
\label{TvsPinj}
\end{figure}

We now consider the dependence of the effective temperature on the mean injected power $\langle \mathcal{P}_\mathrm{inj} \rangle$. This quantity is estimated from the measured velocities $\dot{\xi}_i$:
\begin{equation}
\langle \mathcal{P}_\mathrm{inj} \rangle \sim \rho S_\mathrm{w.m.} \langle \vert \dot{\xi}_1 \vert ^3 + \vert \dot{\xi}_2 \vert ^3 + \vert \dot{\xi}_3 \vert ^3  \rangle
\end{equation}
where $S_\mathrm{w.m.} = 12.10^{-4}~\mathrm{m}^2$ is the submerged section of each wave-maker. It should not be confused with the mean energy flux though scales $\epsilon$ of wave turbulence theory, as most of the injected energy drives bulk flows \cite{Deike2014}. A careful experimental study of the relationship between these two quantities has evidenced the scaling $\langle \mathcal{P}_\mathrm{inj} \rangle \propto \rho \mathcal{S} \sqrt{\epsilon }$ (see \cite{Deike2014}), and \eqref{temperature_vs_p} thus predicts $k_\mathrm{B}T \propto \langle \mathcal{P}_\mathrm{inj} \rangle$. Experimental data are reported in Fig. \ref{TvsPinj} and follow a linear law $T_\mathrm{eff}\propto \langle \mathcal{P}_\mathrm{inj} \rangle$ over three decades.

We finally discuss the role of dissipation. In wave turbulence, damping has been recognized as a source of discrepancy between theory (that assumes a dissipation localized in a restricted frequency range) and experiments, for surface waves \cite{Deike2014} as well as for other wave fields (\textit{e.g.} vibrating plates \cite{Miquel2014}), in which this hypothesis is not fulfilled. To be more precise, energy in weak wave turbulence theory is injected at a given frequency and fully transferred to a dissipative scale (similarly to the Kolmogorov picture of hydrodynamic turbulence), whereas it is essentially dissipated at the injection scale in experiments \cite{Deike2014}. The latter description also applies in this experiment, one reason being that a $f^{-1}$ height spectrum is not steep enough for low-frequency waves to reach a sizable amplitude, hence to dissipate a significant part of the injected energy. This can be seen from the dissipation spectrum $D_{\eta}(f)$, that characterizes the amount of energy dissipated by surface waves of frequency $f$ \cite{Deike2014} : $S_\eta(f) \propto f^{-1}$ leads to $D_\eta(f) \propto f^{5/3}$ or $D_\eta(f) \propto f^{3/2}$ depending on the source of dissipation considered (respectively viscosity in the bulk and surface contamination). Dissipating energy at the injection scales does not prevent the observation of the equilibrium range studied here: in contrast with an energy cascade whose aim is to eventually dissipate energy, the existence of the equilibrium range relies on the hypothesis that no energy is dissipated at these frequencies. It also constrains the setup for such observations: experiments have to be carried out in small vessels, with fluids of low kinematic viscosities and with a large number of efficient resonant interactions at injection scales, otherwise bidirectionnal energy cascades would be generated \cite{Abdurakhimov2015}. For instance, with our setup, we could not observe thermal equilibrium states if water was used instead of mecury, as well as if the forcing was narrow-band, monochromatic, too low or if $f_\mathrm{inj}>100 \mathrm{Hz}$.

\paragraph{Conclusion.---} We experimentally evidenced a thermal equilibrium state over an out-of-equilibrium background in capillary wave turbulence. Its power spectrum density can be used to define an effective temperature that is strongly linked with the statistical properties of this range of scales (\textit{e.g.} shape of the PDF) and to the ones of the out-of-equilibrium state (\textit{e.g.} energy flux). The temperatures obtained are more than ten order of magnitudes higher than the room temperature and we emphasize that such equilibrium states display non-linear phenomena, such as strong coupling between modes.
The precise conditions requiered for the observation of such equilibrium state, as well as the role of the transition to gravity waves remain open questions.

Being able to characterize some scales of an out-of-equilibrium system by equilibrium statistics seems promising. In particular, one may wonder
if other tools of equilibrium statistical mechanics can be used. For
instance the equation of state, the response coefficients and the
fluctuation-dissipation relations could be investigated and compared to
their equilibrium counterparts.

%We also pointed out that dissipation has to be taken into account in the forcing scales in order to get a full picture of the wave-field dynamics.

This work is supported by CNES and ANR-12-BS04-0005-02.


\begin{thebibliography}{10}%
\makeatletter
\providecommand \@ifxundefined [1]{%
 \ifx #1\undefined \expandafter \@firstoftwo
 \else \expandafter \@secondoftwo
\fi
}%
\providecommand \@ifnum [1]{%
 \ifnum #1\expandafter \@firstoftwo
 \else \expandafter \@secondoftwo
\fi
}%
\providecommand \enquote [1]{``#1''}%
\providecommand \bibnamefont  [1]{#1}%
\providecommand \bibfnamefont [1]{#1}%
\providecommand \citenamefont [1]{#1}%
\providecommand\href[0]{\@sanitize\@href}%
\providecommand\@href[1]{\endgroup\@@startlink{#1}\endgroup\@@href}%
\providecommand\@@href[1]{#1\@@endlink}%
\providecommand \@sanitize [0]{\begingroup\catcode`\&12\catcode`\#12\relax}%
\@ifxundefined \pdfoutput {\@firstoftwo}{%
 \@ifnum{\z@=\pdfoutput}{\@firstoftwo}{\@secondoftwo}%
}{%
 \providecommand\@@startlink[1]{\leavevmode}%
 \providecommand\@@endlink[0]{}%
}{%
 \providecommand\@@startlink[1]{%
  \leavevmode
  \pdfstartlink
   attr{/Border[0 0 1 ]/H/I/C[0 1 1]}%
   user{/Subtype/Link/A<</Type/Action/S/URI/URI(#1)>>}%
  \relax
 }%
 \providecommand\@@endlink[0]{\pdfendlink}%
}%
\providecommand \url  [0]{\begingroup\@sanitize \@url }%
\providecommand \@url [1]{\endgroup\@href {#1}{\urlprefix}}%
\providecommand \urlprefix [0]{URL }%
\providecommand \Eprint[0]{\href }%
\@ifxundefined \urlstyle {%
  \providecommand \doi [1]{doi:\discretionary{}{}{}#1}%
}{%
  \providecommand \doi [0]{doi:\discretionary{}{}{}\begingroup
  \urlstyle{rm}\Url }%
}%
\providecommand \doibase [0]{http://dx.doi.org/}%
\providecommand \Doi[1]{\href{\doibase#1}}%
\providecommand \bibAnnote [3]{%
  \BibitemShut{#1}%
  \begin{quotation}\noindent
    \textsc{Key:}\ #2\\\textsc{Annotation:}\ #3%
  \end{quotation}%
}%
\providecommand \bibAnnoteFile [2]{%
  \IfFileExists{#2}{\bibAnnote {#1} {#2} {\input{#2}}}{}%
}%
\providecommand \typeout [0]{\immediate \write \m@ne }%
\providecommand \selectlanguage [0]{\@gobble}%
\providecommand \bibinfo [0]{\@secondoftwo}%
\providecommand \bibfield [0]{\@secondoftwo}%
\providecommand \translation [1]{[#1]}%
\providecommand \BibitemOpen[0]{}%
\providecommand \bibitemStop [0]{}%
\providecommand \bibitemNoStop [0]{.\EOS\space}%
\providecommand \EOS [0]{\spacefactor3000\relax}%
\providecommand \BibitemShut [1]{\csname bibitem#1\endcsname}%
%</preamble>
\bibitem{Herault2015}%
  \BibitemOpen
  \bibfield{author}{%
  \bibinfo {author} {\bibfnamefont{J.}\ \bibnamefont{Herault}},\  \bibinfo
  {author} {\bibfnamefont{F.}\ \bibnamefont{P\'etr\'elis}}\ and\ \bibinfo
  {author} {\bibfnamefont{S.}\ \bibnamefont{Fauve}},\ }%
  \bibfield{journal}{%
  \bibinfo {journal} {EPL}\ }%
  \textbf{\bibinfo {volume} {111}},\ \bibinfo {pages} {44002} (\bibinfo {year}
  {2015}).%
 % \bibAnnoteFile{NoStop}{Eckart}%

\bibitem{Ravelet2008}%
  \BibitemOpen
  \bibfield{author}{%
  \bibinfo {author} {\bibfnamefont{F.}\ \bibnamefont{Ravelet}},\  \bibinfo
  {author} {\bibfnamefont{A.}\ \bibnamefont{Chiffaudel}}\ and\ \bibinfo
  {author} {\bibfnamefont{F.}\ \bibnamefont{Daviaud}},\ }%
  \bibfield{journal}{%
  \bibinfo {journal} {J. Fluid. Mech.}\ }%
  \textbf{\bibinfo {volume} {601}},\ \bibinfo {pages} {339} (\bibinfo {year}
  {2008}).%

\bibitem{Frisch1995}%
  \BibitemOpen
  \bibfield{author}{%
  \bibinfo {author} {\bibfnamefont{U.}\ \bibnamefont{Frish}},\ }%
  \bibfield{journal}{%
  \bibinfo {journal} {Cambridge University Press, Turbulence: The Legacy of A. N. Kolmogorov (Cambridge, England)}\ }%
  (\bibinfo {year}
  {1963}).%
%  \bibAnnoteFile{NoStop}{Liebermann}%


  
  \bibitem{Dallas2015}%
  \BibitemOpen
  \bibfield{author}{%
  \bibinfo {author} {\bibfnamefont{V.}\ \bibnamefont{Dallas}},\  \bibinfo
  {author} {\bibfnamefont{S.}\ \bibnamefont{Fauve}}\ and\ \bibinfo
  {author} {\bibfnamefont{A.}\ \bibnamefont{Alexakis}},\ }%
  \bibfield{journal}{%
  \bibinfo {journal} {Phys. Rev. Lett.}\ }%
  \textbf{\bibinfo {volume} {115}},\ \bibinfo {pages} {204501} (\bibinfo {year}
  {2015}).%
 % \bibAnnoteFile{NoStop}{Eckart}%


\bibitem{Nazarenko2011}%
  \BibitemOpen
  \bibfield{author}{%
  \bibinfo {author} {\bibfnamefont{S.}\ \bibnamefont{Nazarenko}},\ }%
  \bibfield{journal}{%
  \bibinfo {journal} {Wave Turbulence}\ }%
  (\bibinfo {year}
  {2011}).%

  \bibitem{Falcon2007}%
  \BibitemOpen
  \bibfield{author}{%
  \bibinfo {author} {\bibfnamefont{E.}\ \bibnamefont{Falcon}},\  \bibinfo
  {author} {\bibfnamefont{C.}\ \bibnamefont{Laroche}}\ and\ \bibinfo
  {author} {\bibfnamefont{S.}\ \bibnamefont{Fauve}},\ }%
  \bibfield{journal}{%
  \bibinfo {journal} {Phys. Rev. Lett.}\ }%
  \textbf{\bibinfo {volume} {98}},\ \bibinfo {pages} {094503} (\bibinfo {year}
  {2007}).%
    
     \bibitem{Deike2011}
  \BibitemOpen
  \bibfield{author}{%
  \bibinfo {author} {\bibfnamefont{L.}\ \bibnamefont{Deike}},\  \bibinfo
  {author} {\bibfnamefont{C.}\ \bibnamefont{Laroche}}\ and\ \bibinfo
  {author} {\bibfnamefont{E.}\ \bibnamefont{Falcon}},\ }%
  \bibfield{journal}{%
  \bibinfo {journal} {EPL}\ }%
  \textbf{\bibinfo {volume} {96}},\ \bibinfo {pages} {34004} (\bibinfo {year}
  {2011}).%
    
     \bibitem{Chabchoud2011}
  \BibitemOpen
  \bibfield{author}{%
  \bibinfo {author} {\bibfnamefont{A.}\ \bibnamefont{Chabcoub}},\  \bibinfo
  {author} {\bibfnamefont{N. P.}\ \bibnamefont{Hoffmann}}\ and\ \bibinfo
  {author} {\bibfnamefont{N.}\ \bibnamefont{Akhmediev}},\ }%
  \bibfield{journal}{%
  \bibinfo {journal} {Phys. Rev. Lett.}\ }%
  \textbf{\bibinfo {volume} {106}},\ \bibinfo {pages} {204502} (\bibinfo {year}
  {2011}).%

     \bibitem{Tufillaro1989}
  \BibitemOpen
  \bibfield{author}{%
  \bibinfo {author} {\bibfnamefont{N. B.}\ \bibnamefont{Tufillaro}},\  \bibinfo
  {author} {\bibfnamefont{R.}\ \bibnamefont{Ramshankar}}\ and\ \bibinfo
  {author} {\bibfnamefont{J. P.}\ \bibnamefont{Gollub}},\ }%
  \bibfield{journal}{%
  \bibinfo {journal} {Phys. Rev. Lett.}\ }%
  \textbf{\bibinfo {volume} {62}},\ \bibinfo {pages} {422} (\bibinfo {year}
  {1989}).%

     \bibitem{Balkovsky1995}
  \BibitemOpen
  \bibfield{author}{%
  \bibinfo {author} {\bibfnamefont{E.}\ \bibnamefont{Balkovsky}},\  \bibinfo
  {author} {\bibfnamefont{G.}\ \bibnamefont{Falkovich}},\  \bibinfo
  {author} {\bibfnamefont{V.}\ \bibnamefont{Lebedev}}\ and\ \bibinfo
  {author} {\bibfnamefont{I. Ya.}\ \bibnamefont{Shapiro}},\ }%
  \bibfield{journal}{%
  \bibinfo {journal} {Phys. Rev. E}\ }%
  \textbf{\bibinfo {volume} {52}},\ \bibinfo {pages} {4537} (\bibinfo {year}
  {1995}).%

     \bibitem{VanDorn1966}
  \BibitemOpen
  \bibfield{author}{%
  \bibinfo {author} {\bibfnamefont{W. G.}\ \bibnamefont{Van Dorn}},\ }%
  \bibfield{journal}{%
  \bibinfo {journal} {J. Fluid Mech.}\ }%
  \textbf{\bibinfo {volume} {24}},\ \bibinfo {pages} {769} (\bibinfo {year}
  {1966}).%

     \bibitem{Abdurakhimov2015}
  \BibitemOpen
  \bibfield{author}{%
  \bibinfo {author} {\bibfnamefont{L. V.}\ \bibnamefont{Abdurakhimov}},\ \bibinfo
  {author} {\bibfnamefont{M.}\ \bibnamefont{Arefin}},\ \bibinfo
  {author} {\bibfnamefont{G. V.}\ \bibnamefont{Kolmakov}},\ \bibinfo
  {author} {\bibfnamefont{A. A.}\ \bibnamefont{Levchenko}},\ \bibinfo
  {author} {\bibfnamefont{Yu. V.}\ \bibnamefont{Lvov}}\ and\ \bibinfo
  {author} {\bibfnamefont{I. A.}\ \bibnamefont{Remizov}},\ }%
  \bibfield{journal}{%
  \bibinfo {journal} {Phys. Rev. E}\ }%
  \textbf{\bibinfo {volume} {91}},\ \bibinfo {pages} {023021} (\bibinfo {year}
  {2015}).%

     \bibitem{Lvov2015}
  \BibitemOpen
  \bibfield{author}{%
  \bibinfo {author} {\bibfnamefont{Yu. V.}\ \bibnamefont{Lvov}},\ \bibinfo
  {author} {\bibfnamefont{H.}\ \bibnamefont{Andy}}\ and\ \bibinfo
  {author} {\bibfnamefont{G. V.}\ \bibnamefont{Kolmakov}},\ }%
  \bibfield{journal}{%
  \bibinfo {journal} {EPL}\ }%
  \textbf{\bibinfo {volume} {112}},\ \bibinfo {pages} {24004} (\bibinfo {year}
  {2015}).%

     \bibitem{Katyl1968}
  \BibitemOpen
  \bibfield{author}{%
  \bibinfo {author} {\bibfnamefont{R. H.}\ \bibnamefont{Katyl}}\ and\ \bibinfo
  {author} {\bibfnamefont{U.}\ \bibnamefont{Ingard}},\ }%
  \bibfield{journal}{%
  \bibinfo {journal} {Phys. Rev. Lett.}\ }%
  \textbf{\bibinfo {volume} {20}},\ \bibinfo {pages} {248} (\bibinfo {year}
  {1968}).%

     \bibitem{Bouchiat1969}
  \BibitemOpen
  \bibfield{author}{%
  \bibinfo {author} {\bibfnamefont{M. A.}\ \bibnamefont{Bouchiat}}\ and\ \bibinfo
  {author} {\bibfnamefont{J.}\ \bibnamefont{Meunier}},\ }%
  \bibfield{journal}{%
  \bibinfo {journal} {Phys. Rev. Lett.}\ }%
  \textbf{\bibinfo {volume} {23}},\ \bibinfo {pages} {752} (\bibinfo {year}
  {1969}).%

     \bibitem{Zakharov1967}
  \BibitemOpen
  \bibfield{author}{%
  \bibinfo {author} {\bibfnamefont{V. E.}\ \bibnamefont{Zakharov}}\ and\ \bibinfo
  {author} {\bibfnamefont{N. N.}\ \bibnamefont{Filonenko}},\ }%
  \bibfield{journal}{%
  \bibinfo {journal} {J. Appl. Mech. Tech.}\ }%
  \textbf{\bibinfo {volume} {8}},\ \bibinfo {pages} {37} (\bibinfo {year}
  {1967}).%
  
       \bibitem{Deike2014}
  \BibitemOpen
  \bibfield{author}{%
  \bibinfo {author} {\bibfnamefont{L.}\ \bibnamefont{Deike}},\ \bibinfo
  {author} {\bibfnamefont{M.}\ \bibnamefont{Berhanu}}\ and\ \bibinfo
  {author} {\bibfnamefont{E.}\ \bibnamefont{Falcon}},\ }%
  \bibfield{journal}{%
  \bibinfo {journal} {Phys. Rev. E}\ }%
  \textbf{\bibinfo {volume} {89}},\ \bibinfo {pages} {023003} (\bibinfo {year}
  {2014}).%
  
       \bibitem{Pan2015}
  \BibitemOpen
  \bibfield{author}{%
  \bibinfo {author} {\bibfnamefont{Y.}\ \bibnamefont{Pan}}\ and\ \bibinfo
  {author} {\bibfnamefont{D. K. P.}\ \bibnamefont{Yue}},\ }%
  \bibfield{journal}{%
  \bibinfo {journal} {J. Fluid Mech.}\ }%
  \textbf{\bibinfo {volume} {780}},\ \bibinfo {pages} {R1} (\bibinfo {year}
  {2015}).%
  
     \bibitem{Falcon2009}
  \BibitemOpen
  \bibfield{author}{%
  \bibinfo {author} {\bibfnamefont{C.}\ \bibnamefont{Falc\'on}},\ \bibinfo
  {author} {\bibfnamefont{E.}\ \bibnamefont{Falcon}},\ \bibinfo
  {author} {\bibfnamefont{U.}\ \bibnamefont{Bortolozzo}}\ and\ \bibinfo
  {author} {\bibfnamefont{S.}\ \bibnamefont{Fauve}},\ }%
  \bibfield{journal}{%
  \bibinfo {journal} {EPL}\ }%
  \textbf{\bibinfo {volume} {86}},\ \bibinfo {pages} {14002} (\bibinfo {year}
  {2009}).%

     \bibitem{Falcon2011}
  \BibitemOpen
  \bibfield{author}{%
  \bibinfo {author} {\bibfnamefont{E.}\ \bibnamefont{Falcon}}\ and\ \bibinfo
  {author} {\bibfnamefont{C.}\ \bibnamefont{Laroche}},\ }%
  \bibfield{journal}{%
  \bibinfo {journal} {EPL}\ }%
  \textbf{\bibinfo {volume} {95}},\ \bibinfo {pages} {34003} (\bibinfo {year}
  {2011}).%


     \bibitem{Henry2000}
  \BibitemOpen
  \bibfield{author}{%
  \bibinfo {author} {\bibfnamefont{E.}\ \bibnamefont{Henry}},\ \bibinfo
  {author} {\bibfnamefont{P.}\ \bibnamefont{Alstrom}}\ and\ \bibinfo
  {author} {\bibfnamefont{M. T.}\ \bibnamefont{Levinsen}},\ }%
  \bibfield{journal}{%
  \bibinfo {journal} {EPL}\ }%
  \textbf{\bibinfo {volume} {52}},\ \bibinfo {pages} {27} (\bibinfo {year}
  {2000}).%


     \bibitem{Deike2014B}
  \BibitemOpen
  \bibfield{author}{%
  \bibinfo {author} {\bibfnamefont{L.}\ \bibnamefont{Deike}},\ \bibinfo
  {author} {\bibfnamefont{D.}\ \bibnamefont{Fuster}},\ \bibinfo
  {author} {\bibfnamefont{M.}\ \bibnamefont{Berhanu}}\ and\ \bibinfo
  {author} {\bibfnamefont{E.}\ \bibnamefont{Falcon}},\ }%
  \bibfield{journal}{%
  \bibinfo {journal} {Phys. Rev. Lett.}\ }%
  \textbf{\bibinfo {volume} {112}},\ \bibinfo {pages} {234501} (\bibinfo {year}
  {2014}).%


     \bibitem{Ruban2006}
  \BibitemOpen
  \bibfield{author}{%
  \bibinfo {author} {\bibfnamefont{V. P.}\ \bibnamefont{Ruban}},\ }%
  \bibfield{journal}{%
  \bibinfo {journal} {Phys. Rev. E}\ }%
  \textbf{\bibinfo {volume} {74}},\ \bibinfo {pages} {036305} (\bibinfo {year}
  {2006}).%


     \bibitem{Shats2010}
  \BibitemOpen
  \bibfield{author}{%
  \bibinfo {author} {\bibfnamefont{M.}\ \bibnamefont{Shats}}\ and\ \bibinfo
  {author} {\bibfnamefont{H.}\ \bibnamefont{Punzmann}}\ and\ \bibinfo
  {author} {\bibfnamefont{H.}\ \bibnamefont{Xia}},\ }%
  \bibfield{journal}{%
  \bibinfo {journal} {Phys. Rev. Lett.}\ }%
  \textbf{\bibinfo {volume} {104}},\ \bibinfo {pages} {104503} (\bibinfo {year}
  {2010}).%






     \bibitem{Stokes1847}
  \BibitemOpen
  \bibfield{author}{%
  \bibinfo {author} {\bibfnamefont{G. G.}\ \bibnamefont{Stokes}},\ }%
  \bibfield{journal}{%
  \bibinfo {journal} {Trans. Camb. Philo. Soc.}\ }%
  \textbf{\bibinfo {volume} {8}},\ \bibinfo {pages} {441} (\bibinfo {year}
  {1847}).%

   \bibitem{Crapper1957}
  \BibitemOpen
  \bibfield{author}{%
  \bibinfo {author} {\bibfnamefont{G. D.}\ \bibnamefont{Crapper}},\ }%
  \bibfield{journal}{%
  \bibinfo {journal} {J. Fluid Mech.}\ }%
  \textbf{\bibinfo {volume} {2}},\ \bibinfo {pages} {532} (\bibinfo {year}
  {1957}).%

 

     \bibitem{Miquel2014}
  \BibitemOpen
  \bibfield{author}{%
  \bibinfo {author} {\bibfnamefont{B.}\ \bibnamefont{Miquel}},\ \bibinfo
  {author} {\bibfnamefont{A.}\ \bibnamefont{Alexakis}}\ and\ \bibinfo
  {author} {\bibfnamefont{N.}\ \bibnamefont{Mordant}},\ }%
  \bibfield{journal}{%
  \bibinfo {journal} {Phys. Rev. E}\ }%
  \textbf{\bibinfo {volume} {89}},\ \bibinfo {pages} {062925} (\bibinfo {year}
  {2014}).%




\end{thebibliography}
\end{document}